\documentclass[11pt,twoside]{article}
\usepackage{Cargesepasp}
\usepackage{epsf}
\def\edcomment#1{\iffalse\marginpar{\raggedright\sl#1\/}\else\relax\fi}
\reversemarginpar

\begin{document}
\title{Science with SIRTF -- Some Examples} 
\author{Bernhard Brandl and the IRS Team} 
\affil{Cornell University, 222 Space
  Sciences Building, Ithaca, NY 14853, USA}

\begin{abstract}
  The Space InfraRed Telescope Facility (SIRTF) is a space-borne,
  cryogenically-cooled infrared observatory capable of studying
  objects ranging from our Solar System to the distant reaches of the
  Universe.  SIRTF is the final element in NASA's Great Observatories
  Program, and an important component of its Origins Program.  The
  intent of this paper is to summarize the most important parameters
  of the observatory and its scientific instruments and to present
  three typical examples of scientific areas where SIRTF will have a
  significant impact.  These examples of ``nearby'' targets -- brown
  dwarf surveys, protostellar disks, and massive young clusters --
  have been selected in regard to the main topics of this conference.
\end{abstract}

\section{Introduction}
\begin{figure}
\plotone{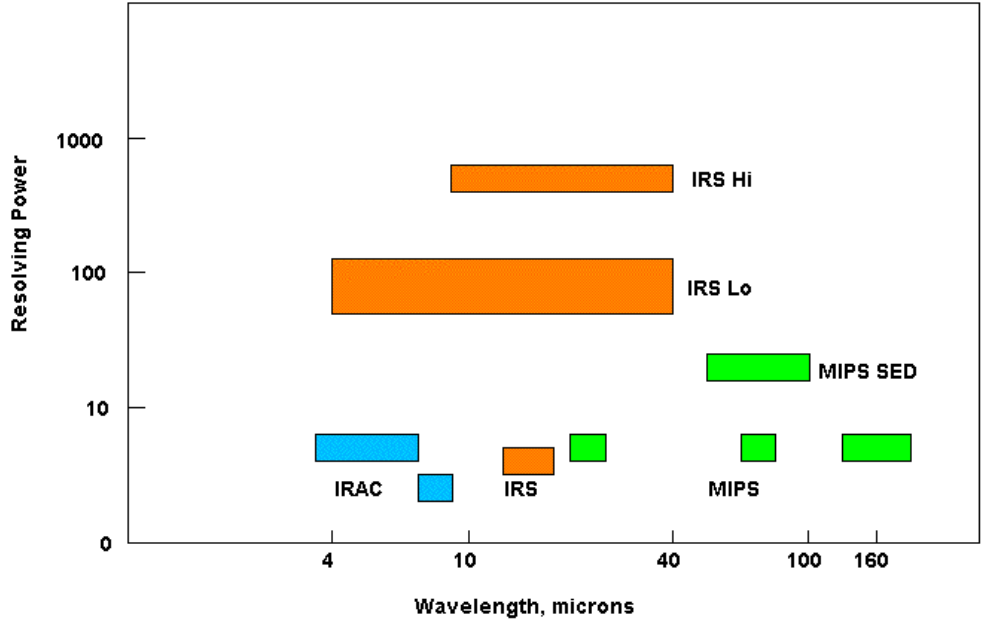}
\caption{\label{lamres} Wavelength versus spectral resolution coverage 
  of the scientific instruments onboard SIRTF.}
\end{figure}

The Space InfraRed Telescope Facility (SIRTF) (Fanson et al. 1998) is
the fourth and final element in NASA's family of ``Great
Observatories''.  SIRTF consists of a 0.85-meter telescope, cooled to
$T < 5.5K$ and three cryogenically-cooled science instruments capable
of performing imaging and spectroscopy in the $3 - 180 \mu$m
wavelength range (Fig.~\ref{lamres}).  More details are summarized in
Table~\ref{sirtffacts}.

\begin{table}\centering
\begin{tabular}{| l l |}
\hline
Launch Date & December 2001 \\
Launch Vehicule/Site & Delta 7920H ELV / Kennedy Space Center \\
Estimated Lifetime & 2.5 years (minimum); 5+ years (goal) \\
Orbit & Earth-trailing, Heliocentric \\
Wavelength Coverage & 3 - 180 microns \\
Telescope & 85 cm diameter, f/12 lightweight Beryllium \\
Diffraction Limit & 6.5 microns \\
Science Capabilities & Imaging / Photometry, 3-180 microns \\
                     & Spectroscopy, 5-40 microns \\
                     & Spectrophotometry, 50-100 microns \\
Planetary Tracking & 1 arcsec / sec \\
Cryogen / Volume & Liquid Helium / 360 liters \\
Launch Mass & 950 kg \\ \hline
\end{tabular}
\caption{\label{sirtffacts}Summary of important SIRTF parameters}
\end{table}

Incorporating large-format infrared detector arrays, SIRTF offers
orders-of-magnitude improvements in capability over existing programs.
While SIRTF's mission lifetime requirement remains 2.5 years, recent
programmatic and engineering developments have brought a 5-year
cryogenic mission within reach. A fast-track development schedule will
lead to a launch in December 2001. SIRTF represents an important
scientific and technical bridge to NASA's new Origins program.

75\% of the observing time will be available to the community in form
of general science proposals and so-called Legacy proposals (The SIRTF
Legacy Science Program is intended to maximize the scientific return
from SIRTF early in the mission, and comprises a handful of
large-scale observing projects to be executed primarily within the
first year of SIRTF's lifetime.)  In addition for the first 2.5 years
of SIRTF's prime mission, 20 percent of observing time will be
allocated to the GTO program and an additional 5\% is reserved as
Director's discretionary time.
More details can be found at {\tt http://sirtf.caltech.edu}

\begin{figure}
\plotone{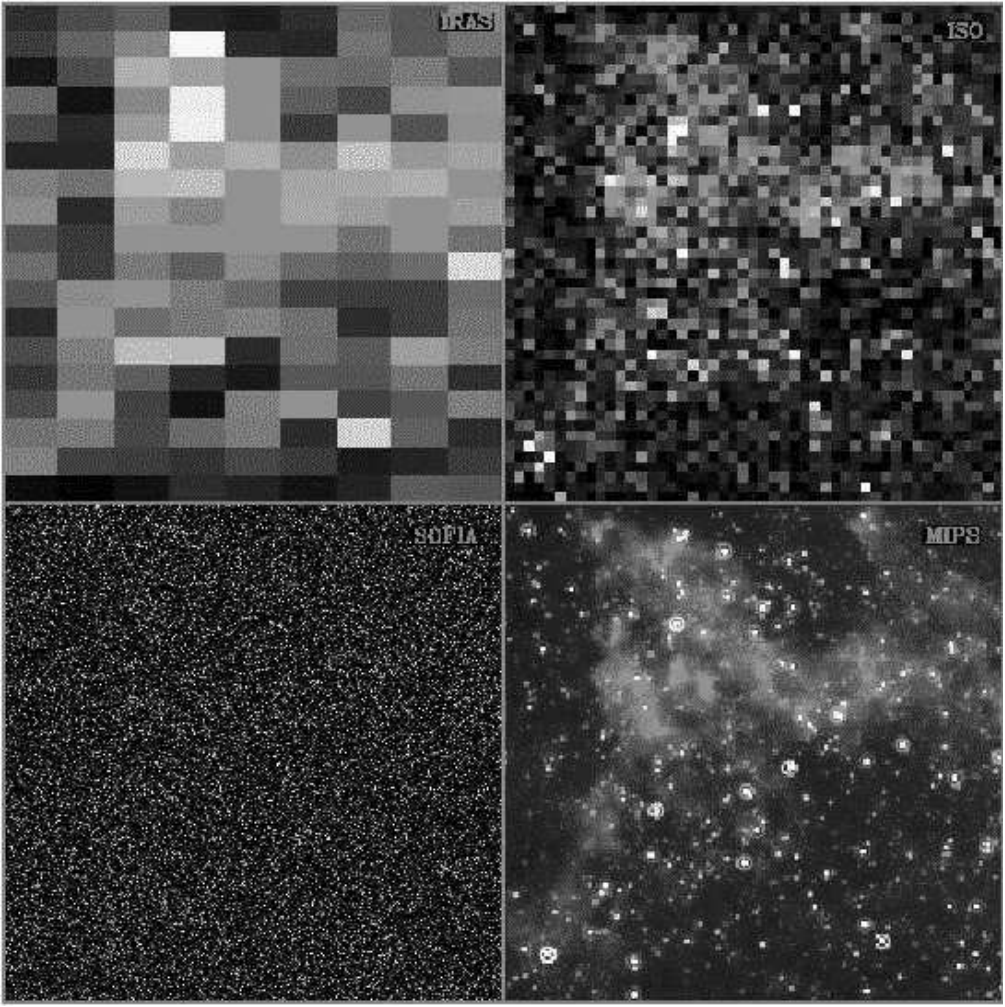}
\caption{\label{instcomp} These are logarithmically scaled versions of 
  comparison images generated for IRAS, ISO, SOFIA, and MIPS. A test
  field was "observed" with these four instruments, giving each
  instrument 24 hours of integration time and taking into account the
  sensitivity, the plate scale, and the field of view. The IRAS image
  has very large pixels and is really only capable of detecting the
  infrared cirrus in this field. The ISO image has better spatial
  resolution but is limited by the small field-of-view and low
  sensitivity of the arrays. SOFIA has excellent spatial resolution
  because of the large (2.5m) telescope but a correspondingly small
  field-of-view (even with a 32x32 array) and is limited in
  sensitivity because it uses warm optics. The predicted MIPS
  performance on the test field is excellent because of the high
  sensitivity of the detectors, good spatial resolution, and the large
  field-of-view of the 32x32 array.  (Engelbracht et al. 1999).}
\end{figure}

This paper presents three examples of ``typical'' SIRTF science, most
of which are covered by the GTO program.  These examples have been
selected to demonstrate SIRTF's superb capabilities and to match the
scientific focus of the 3rd ``Three Island Euroconference on Clusters
and Associations''.  More examples can be found elsewhere, e.g., in
Brandl et al. (1999).

\section{Instruments and Sensitivities}
SIRTF's science payload consists of three cryogenically cooled
instruments, which together offer observational capabilities
stretching from the near- to the far-infrared. Figure~\ref{instcomp}
shows a simulation comparing the performance of the latest major IR
observatories on deep large area surveys. Following is a brief summary
of the characteristics of each instrument.  Details are described in
the SIRTF Observers Manual (2000).

\subsection{The InfraRed Array Camera (IRAC)}
The IRAC (Giovanni G. Fazio, PI) provides images at 3.6, 4.5, 5.8 and
8.0$\mu$m, over two adjacent 5.12 x 5.12 arcminute fields of view. One
field of view images simultaneously at 3.6 and 5.8$\mu$m and the other
at 4.5 and 8.0$\mu$m via a dichroic beamsplitter. All four detector
arrays are 256x256 with 1.2 arcsecond square pixels. A cold shutter is
provided for dark measurements, and also allows the detectors to be
illuminated by light from a transmission calibrator. Sensitivities for
IRAC are given in Table~\ref{iracsens}.

\begin{table}\centering
\begin{tabular}{| c c c |}
\hline
waveband [$\mu$m] & sensitivity [$\mu$Jy]   & saturation [Jy] \\
(center wavelength) & (1$\sigma$ in 100s) & (0.4s, point source) \\ \hline
3.6  & 1.29 & 1.39 \\
4.5  & 1.86 & 1.19 \\
5.8 & 6.09 & 2.05 \\
8.0 & 7.97 & 1.35 \\ \hline
\end{tabular}
\caption{\label{iracsens} IRAC point source sensitivities and saturation 
flux densities}
\end{table}

\subsection{The InfraRed Spectrograph(IRS)}
The IRS (James R. Houck, PI) performs both low and high-resolution
spectroscopy. Low resolution, long slit spectra
($\lambda/\Delta\lambda = 62 - 124$) can be obtained from 5.3 to
40$\mu$m. High resolution spectra ($\lambda/\Delta\lambda = 600$) in
Echelle mode can be obtained from 10 to 37$\mu$m. The spectrograph
consists of four modules, each of which is built around a 128x128
pixel array.  The sizes of the slits are summarized in
table~\ref{irsslits}. 

\begin{table}\centering
\begin{tabular}{| c c c c c |}
\hline
wavelength  & resolution & pixel size & slit width & slit length \\
range [$\mu$m] & [$\lambda / \Delta\lambda$] & [$''$] & [$''$] 
& [$''$] \\ \hline
5.3 -- 8.5       & 62 -- 124  & 1.8        &  3.6        &  54.6 \\
7.5 -- 14.2      & 62 -- 124  & 1.8        &  3.6        &  54.6 \\
14.2 -- 21.8     & 62 -- 124  & 4.8        &  9.7        &  151.3 \\
20.6 -- 40.0     & 62 -- 124  & 4.8        &  9.7        &  151.3 \\
10.0 -- 19.5     & 600        & 2.4        &  5.3        &  11.8 \\
19.3 -- 37.0     & 600        & 4.8        &  11.1       &  22.4 \\ \hline
\end{tabular}
\caption{\label{irsslits} Slit sizes of the different IRS modules}
\end{table}

\begin{figure}
\plotone{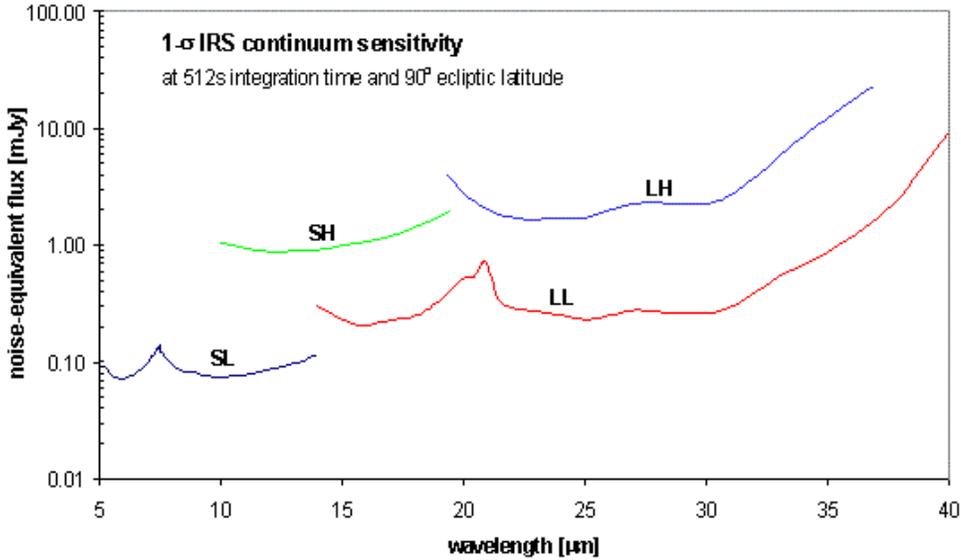}
\caption{\label{irssens} 1$\sigma$ continuum sensitivity of all four IRS 
  modules in a 512 second integration time and 90\deg ecliptic
  latitude as a function of wavelength.}
\end{figure}

One of the modules incorporates a peak-up function that can be used in
locating and positioning sources on any of the four spectrometer slits
with sub-arcsecond precision.  The peak-up array has 1.8 arcsec square
pixels and a field of view of 1 arcminute by 1.2 arcminutes. Two
filters are available for use with up array, covering 13.5 -- 18.5
microns or 18.5 -- 26 microns.  Sensitivities for the IRS are shown in
Figure~\ref{irssens}.

\subsection{The Multiband Imaging Photometer for SIRTF (MIPS)}
The MIPS (George H. Rieke, PI) is designed to provide photometry and
super resolution imaging as well as high speed mapping capabilities in
three wavelength bands centered near 24, 70 and 160 microns. The array
materials are Si:As, Ge:Ga, and stressed Ge:Ga with pixel sizes of
2.5, 5, 10, and 16 arcseconds, depending on wavelength and detector.
MIPS is also capable of low-resolution spectroscopy
($\lambda/\Delta\lambda = 15 - 25$) over the wavelength range 55-96
microns and a Total Power Mode for measuring absolute sky brightness.
Sensitivities for MIPS are given in Table~\ref{mipssens}.

\begin{table}\centering
\begin{tabular}{| c c c |}
\hline
waveband [$\mu$m] & sensitivity [mJy]   & saturation [Jy] \\
(center wavelength)  & (5$\sigma$ in 500s) & (1s, point source) \\ \hline
24  & 0.35 & 13 \\
70  & 1.30 &  9 \\
SED & 6.50 & 45 \\
160 & 22.5 & 40 \\ \hline
\end{tabular}
\caption{\label{mipssens}MIPS point source sensitivities and saturation 
flux densities}
\end{table}

\section{Science with SIRTF}
\subsection{Example 1: Searches for Brown Dwarfs}
During the past two years, ground-based surveys have begun to succeed
in identifying large numbers of relatively young, nearby brown dwarfs
in star-forming regions, young open clusters and the eld. IRAC will be
able to extend those e orts to older and lower mass brown dwarfs and
superplanets. In nearby star-forming regions like Taurus, IRAC should
be able to detect isolated objects down to near the mass of Jupiter,
$~1M_J$.  In the nearest and best-studied open clusters -- in
particular, the Pleiades and the Hyades -- IRAC should be able to
detect brown dwarfs down to $10 M_J$. 

\begin{figure}
\plotone{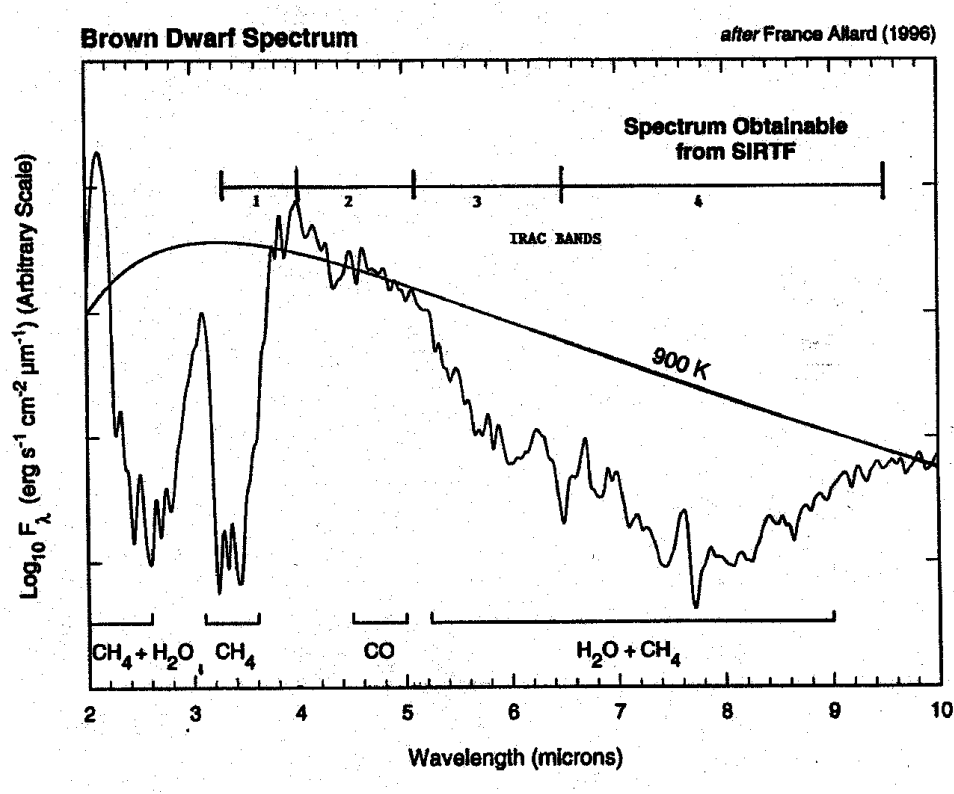}
\caption{\label{bdiracbands} The IRAC bands are shown in relation to a
  model brown dwarf spectrum (Allard et al. 1996) and a 900K blackbody.}
\end{figure}

As Figure~\ref{bdiracbands} illustrates, the ratio of the IRAC bands
at 3.6 and 4.5$\mu$m is a very good means of selecting brown dwarfs.
The method of observation would be to carry out a large area survey of
several open star clusters (e.g. NGC~2264, Pleiades, Hyades). The
survey area is approximately 5 square degrees for NGC~2264 and the
Pleiades, and approximately 10 square degrees for the Hyades.  The
integration time would be 30 seconds per position, observing each
position three times (90s total), which would achieve a sensitivity of
10$\mu$Jy, 5$\sigma$ at 4.5$\mu$m. The total observing time for all
three fields would be about 160 hours. The brown dwarf candidates
would be selected by 3.6/4.5$\mu$m color, based on model predictions
and known objects from the 2MASS and DENIS surveys. The lowest mass
objects will be detected only at 4.5$\mu$m in 90s. The 4.5$\mu$m only
objects would need to be re-observed with much longer integrations to
provide a detection at 3.6$\mu$m. 

\begin{figure}
\plotone{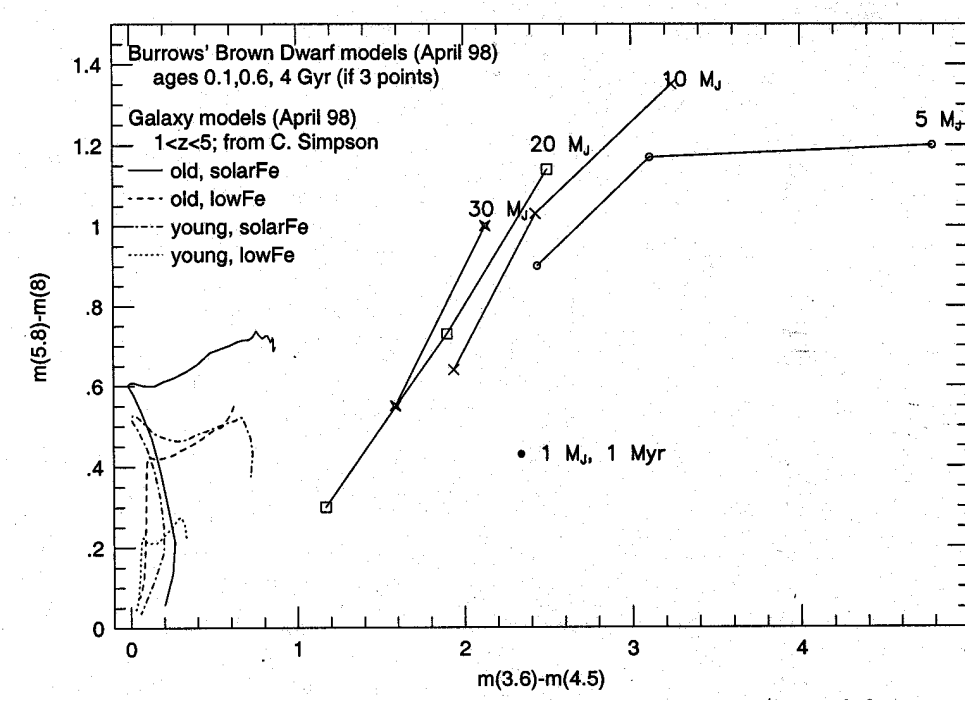}
\caption{\label{bdiractracks} IRAC color/color diagram showing galaxies 
  (lower left corner) and brown dwarfs of different masses and ages
  (upper right). (Fazio et al. 1999 and references therein).}
\end{figure}

Figure \ref{bdiractracks} shows the IRAC color/color diagram for
objects with masses of 5, 10, 20, and $30 M_J$ at various ages.  The
IRAC color/color diagram for galaxies in the field is shown in the
same diagram. The galaxies are well-separated from the brown dwarfs,
appearing in the lower left corner of the color/color plot.  (Fazio et
al. 1999).

\subsection{Example 2: Protostellar and Protoplanetary Disks}
\begin{figure}
\plottwo{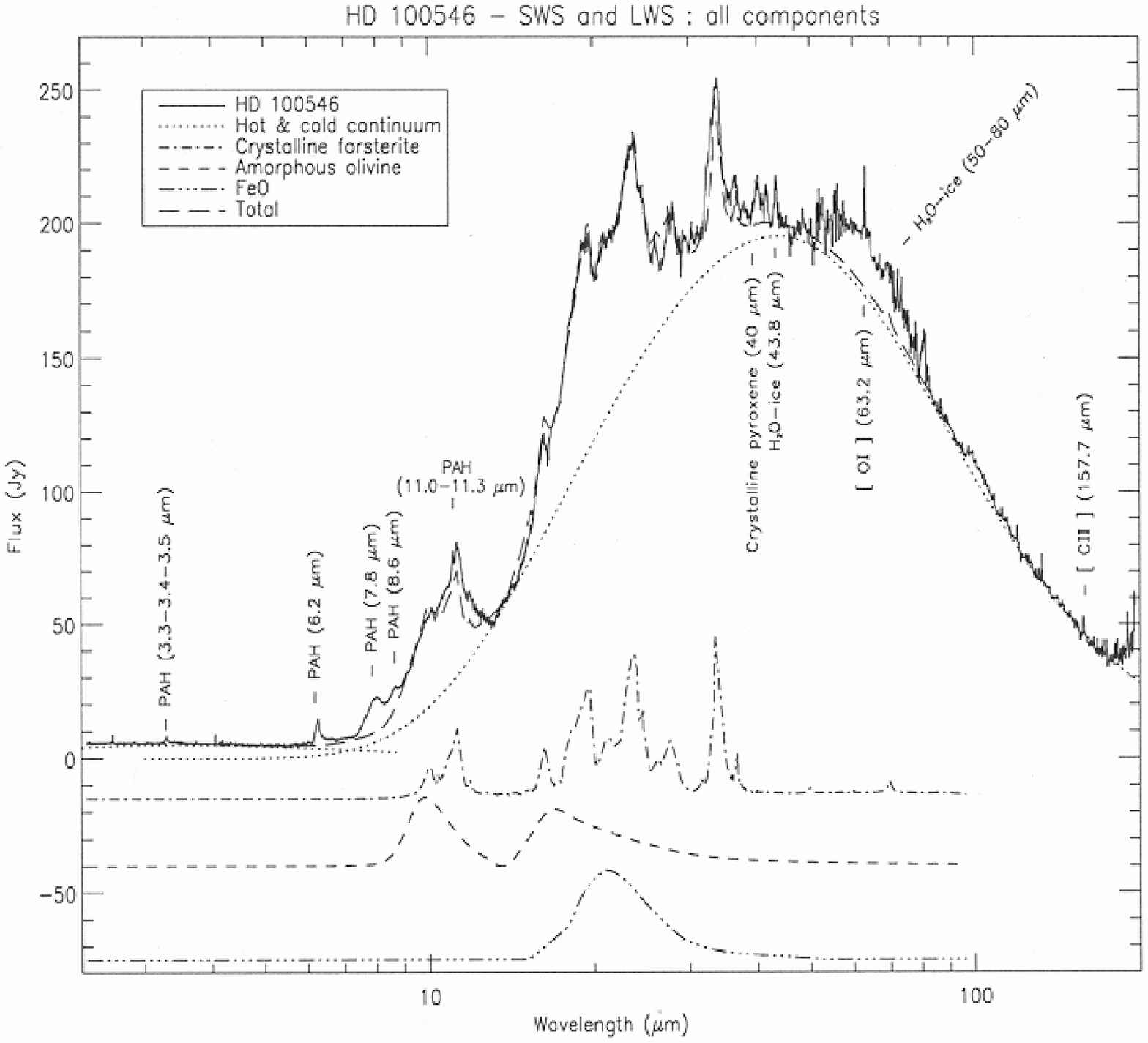}{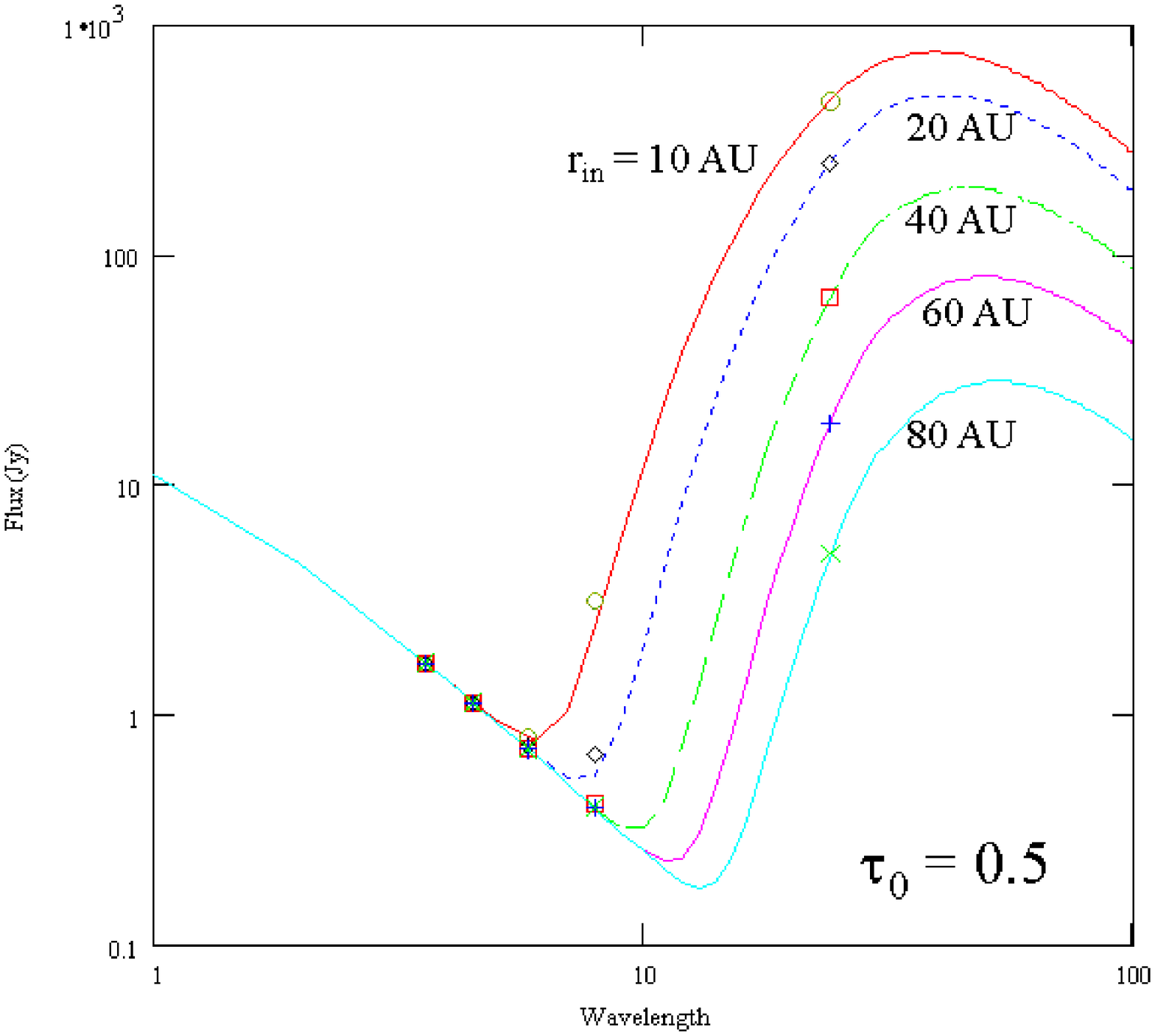}
\caption{\label{hd100546} Left: ISO-SWS/LWS spectrum of the Herbig Ae 
  star HD~100546 (Malfait et al. 1998). Note the wealth of spectral
  features including PAHs (polycyclic aromatic hydrocarbons),
  amorphous and crystalline silicates, rust, etc.  The continuum
  cutoff marks the disk's inner edge.  Right: Model debris disks made
  with HR4796A-like parameters ($L = 18.1 L_\odot$, $T_\star = 9000K$,
  $D = 67$pc, $r_{outer} = 100$ AU).  The plot symbols indicate
  positions of the IRAC and MIPS 24$\mu$m bands.  The further out the
  inner radius of the disk the fainter its radiation relative to the
  stellar continuum at shorter wavelengths.}
\end{figure}

The spectral range covered by the IRS includes most of the important
emission and absorption features from interstellar and
solar-system-like dust grains (Fig.~\ref{hd100546}).  It also includes
many fine structure lines of ions and neutral atoms that are bright in
H\,II regions, phot-dissociation regions and young stellar objects;
the lower transitions in pure rotational spectrum of molecular
hydrogen; and many other molecular rotational and rotation-vibration
lines of abundant species whose emission is important in the
energetics or diagnosis of warm molecular gas.

The IRS is particularly sensitive to emission from dust and gas that
lies in the inner parts of protostellar, protoplanetary and debris
disks, between a few tenths and a few tens of AU from the star.  In
the nearest regions of star formation, the IRS will be able -- in just
a few minutes of integration -- to take high-quality spectra of disks
around protostellar objects with masses down to the hydrogen-burning
limit and even beyond.  Fig.~\ref{hd100546} also illustrates how the
mid-IR SEDs varies under different boundary conditions and hence
can be used as powerful diagnostics for disk evolution and structure,
even if the source is spatially unresolved (Armus et al. 2000).

\subsection{Example 3: Star Formation in Interacting Galaxies}
Collisions between galaxies are one of the most energetic events in
the Universe with energies of typically $10^{53}J$ -- the equivalent
of $~10^{8-9}$ supernovae -- on timescales of about $3\times 10^8$ yrs
(Struck 1999).  The interactions trigger the formation of massive star
clusters either directly through the collision of clouds or induce
bars which funnel gas and foster wave and resonant ring star
formation.  The strongest star formation rates are usually observed in
the most violently interacting galaxies -- the best known example
probably is the Antennae galaxy NGC~4038/39.

The young stars within a starburst are extremely enshrouded by dust
with typical average extinctions of $A_V ~ 25$mag (Smith et al. 1996).
Hence, observational studies in the mid-IR are the best method to
probe the properties of dusty starburst cores and the surrounding
interstellar medium. Fig.~\ref{irsiags1} illustrates the spatial
resolution of the IRS spectrographs with their slits overlayed on the
HST/WFPC2 picture of the Antennae galaxy (Whitmore et al. 1999). At
the distance of 2.8\,kpc the width of the short-high slit
($10-21\mu$m) corresponds to only 500\,pc; the length of the short-
and long-low slits of $55''$ and $151''$, respectively, make it -- for
the first time -- practically possibly to obtain mid-IR spectra of the
entire region at relatively high spatial resolution.

\begin{figure}[ht]
\vspace{10mm}
\caption{\label{irsiags1} The sizes of the IRS short-low (top) and 
  short-high (bottom right) slits projected onto NGC 4038/39. The
  locations and angles of the slits were arbitrarily chosen.} 
\end{figure}

\begin{figure}[ht]
\plotone{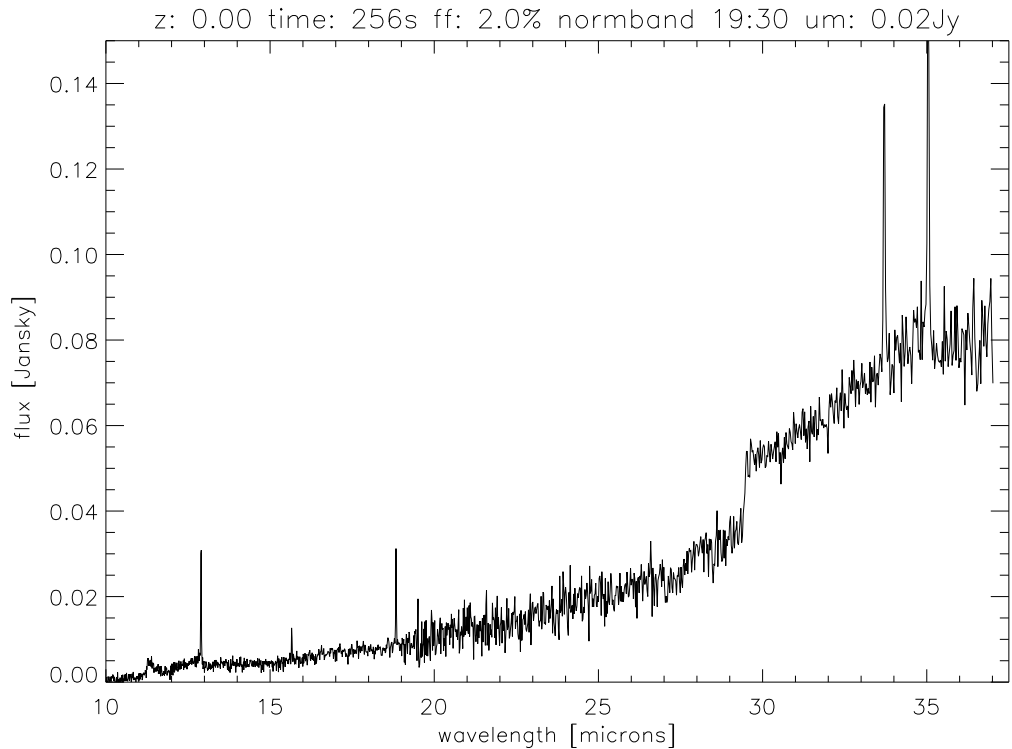}
\caption{\label{irsiags2} Simulated IRS spectrum of a 20mJy source in 
  256s integration time per slit, assuming a M82 like spectrum.  The
  most prominent emission lines are: NeII (12.79$\mu$m), NeIII
  (15.55$\mu$m), SIII (18.71$\mu$m), SIII (33.48$\mu$m), SiII
  (34.80$\mu$m).  The ``jump'' around 30$\mu$m is due to an artifact
  in the input spectrum.}
\end{figure}

Figure~\ref{irsiags2} shows a simulation of how the spectrum of a
low luminosity region in the Antennae galaxy will look when observed
with the IRS high-resolution module.  The on-source integration time
for this simulation is 256s, the flux is normalized to an average
20mJy in the $19-30\mu$m window, a typical value for a low-luminosity
region according to the ISOCAM-CVF map (Mirabel et al. 1998).  The
instrument simulation is based on the mid-IR spectrum of the starburst
galaxy M82 as observed by ISO-SWS (Sturm et al. 2000).  Despite the
relative faintness and short integration time one can see a variety of
important diagnostic lines (see figure caption).

\acknowledgments{This work was supported under JPL contract 960803.}

\end{document}